%% file: turbo_tcomm.tex
\documentclass[twoside]{IEEEtran}

\usepackage[dvips]{graphicx}
\usepackage[tight]{subfigure}
\usepackage[cmex10]{amsmath}
\usepackage{amssymb}

\def \bm {\mathbf}
\newcommand{\scsi}{source coding with side information}

\begin{document}
\title{A Unified Perspective on Parity- and Syndrome-Based Binary Data Compression Using Off-the-Shelf Turbo Codecs}
\author{Lorenzo~Cappellari,~\IEEEmembership{Member,~IEEE,}
        and Andrea~De~Giusti\thanks{Manuscript submitted August 2, 2010.}
\thanks{L.~Cappellari is with the Dept.~of Information Engineering of the University of Padova, Italy. A.~De~Giusti is with Nidek Technologies Srl, Albignasego, PD, Italy.}}

\maketitle

\begin{abstract}
\input{turbo_tcomm.txt}
\end{abstract}


\IEEEpeerreviewmaketitle

\section{Introduction}

\IEEEPARstart{T}{he possibility} of employing turbo codes for data compression has been exploited after that the relation of channel coding with \emph{distributed source coding} was made clear.

The concept of \emph{distributed source coding} (DSC) applies to the scenario where many correlated sources must be compressed \emph{without} allowing the respective encoders, that send their compressed outputs to a common decoder, communicate with each other. In their landmark paper, \cite{slepian_noiselessCodingCorrSrc}, Slepian and Wolf extended to this scenario the well-known result of Shannon for a single information sequence, namely $R\geq H(X)$ for faithful reproduction \cite{shannon1948,cover_ElemInfoThNew}, and showed that in terms of aggregate rate of the compressed representation \emph{there are no losses} with respect to the traditional case where the encoders communicate with each other. In practice, the amount of power used for inter-encoder communications can be saved while achieving the same compression performance. In \cite{wyner_RDFunctSrcCodingSideInfo}, Wyner and Ziv investigated instead the rate-distortion function in the DSC-related scenario of \emph{\scsi} (SCSI) at the decoder, where a source must be compressed as usual but the decoder can also rely on some correlated \emph{side information} for reconstruction within a given distortion limit. Again, if perfect reconstruction is desired, they showed that \emph{there are no losses} with respect to the case where the encoder can access the side information too.

Since the results in \cite{slepian_noiselessCodingCorrSrc} and \cite{wyner_RDFunctSrcCodingSideInfo} are obtained with a non-constructive \emph{random binning} approach, it took about thirty years for practical SCSI systems to appear. As foreseen by Wyner in his 1974 paper \cite{wyner_resultsShannon}, in order to achieve the theoretical limits all of them are based on concepts that are rather typical of \emph{channel coding}, such as \emph{syndromes} or \emph{parities}.

In particular, the very first practical SCSI system appeared in \cite{pradhan_DISCUS1} (see also \cite{pradhan_DISCUS2,pradhan_genCosetCodesBinning}). In this system, the \emph{syndrome} relative to a \emph{trellis code} \cite{forney_CosetI} is computed at the encoder in order to signal the \emph{coset} to which the current (quantized) source outcome belongs. Then, the decoder reconstructs the data relying as well on the side information. 
Similarly, in the case of near-lossless \emph{binary} data compression with \emph{binary} side information, many authors applied the \emph{syndrome-based} approach of \cite{wyner_resultsShannon} relying on \emph{low-density parity check} (LDPC) \cite{mackay_ldpc} or \emph{turbo} \cite{berrou_turboJCOM} \emph{codes}. For example, syndromes relative to LDPC codes are used in \cite{liveris_sideLDPC,schonberg00_DistributedCodeConstructions}, while syndromes relative to turbo codes are used in \cite{liveris_DistributedCompressionOf,tu_SF-ISF,roumy_RateAdaptiveTurbo,zamani09_AFlexibleRate,stankovic06_OnCodeDesign}. While in the case of LDPC codes the syndrome formation is straightforward (due to the fact that LDPC codes are exactly defined by means of their \emph{parity-check} matrix), turbo-code-syndrome formation is less direct. In \cite{liveris_DistributedCompressionOf}, in addition to the \emph{principal trellis} employed in traditional channel coding, \emph{complementary trellises} are used for syndrome formation and decoding (as in \cite{pradhan_DISCUS1}). A specific \emph{parity-check} matrix is instead employed for syndrome formation in \cite{tu_SF-ISF}, \cite{roumy_RateAdaptiveTurbo} and \cite{zamani09_AFlexibleRate}. Decoding is performed by means of standard turbo decoding in \cite{tu_SF-ISF,zamani09_AFlexibleRate} and using the so-called \emph{syndrome trellis} in \cite{roumy_RateAdaptiveTurbo}. Some of these approaches are not limited to the SCSI problem, but can be also applied to DSC, i.e.~where no variable is exactly known at the decoder \cite{schonberg00_DistributedCodeConstructions,stankovic06_OnCodeDesign}.

Formerly, rather than a syndrome-based approach, many binary SCSI-related works dealing in particular with turbo codes took a \emph{parity-based} approach. Within the latter approach, the side information is simply seen as a ``dirty'' version of the source (possibly non-binary) that could be ``channel-decoded'' upon receiving some parity bits, formed by the encoder with respect to a \emph{systematic code}. Even if the syndrome-based approach is provenly optimal while the parity-based one is not always so \cite{tan_SWcodingParitySyndrome}, satisfying results have been reported as well \cite{garcia_compressionTurbo,aaron_compressionSideTurboCodes}. In addition, however, in order to avoid the limitations of the parity-based approaches it is possible to design the parity formation procedure in an optimal way \cite{sartipi08_DistributedSourceCoding}. The parity-based approaches have at least two advantages over syndrome-based ones. First, error-prone encoder-to-decoder transmission over more realistic channels than the traditional \emph{binary symmetric channel} (BSC) or \emph{binary erasure channel} (BEC), over which the parity bits become ``dirty'' (possibly non-binary) parities, can be easily handled. Second, \emph{puncturing} can be immediately used for rate adaptation, and the resulting code is automatically \emph{incremental}. These properties were hence effectively exploited for \emph{joint source-channel coding} of a single information sequence \cite{garcia_CompressionOfBinary,hagenauer_TheTurboPrinciple,Hagenauer2004} or for DSC-based video compression with a feedback channel \cite{girod_DVC}.

In principle, it is possible to puncture turbo-syndrome bits too in order to achieve an \emph{incremental} source code \cite{roumy_RateAdaptiveTurbo,li_AnOptimalDistributed,tan_EnhancingTheRobustness}, but if the parity-check matrix is not properly chosen erasures or flips to the syndrome bits can make the correct reconstruction of some elements of the source very hard and hence lead to a very poor performance \cite{tan_EnhancingTheRobustness}. If syndromes are computed based on LDPC codes, syndrome decoders can instead handle more successfully erased or ``dirty'' syndromes, since decoding based on \emph{message-passing} algorithms on \emph{factor graphs} \cite{kschischang_FactorGraphsAnd} can easily model this scenario. In general, techniques for \emph{syndrome protection} against transmission losses can always be employed that possibly permit to exchange ``soft'' information with an actual syndrome decoder in order to maximize the performance \cite{hu_ANewCoding,varodayan_RateAdaptiveDistributed}. But, many convolutional and turbo syndrome decoders expect a strictly \emph{binary} syndrome as input (and a binary side information), as for example in \cite{liveris_DistributedCompressionOf,tu_SF-ISF}, so that their performance cannot be properly optimized in case of non-binary syndrome transmission channels.

In this paper, we consider the problem of turbo-code-based data compression, with or without side information at the decoder. In particular, we tackle both problems of turbo-parity and turbo-syndrome decoding from the point of view of a general \emph{maximum a posteriori probability} problem. As soon as the probability function to be maximized is factorized into its building terms, it becomes immediately straightforward to understand how conventional and \emph{ready-available} iterative decoding algorithms used for turbo decoding can be applied to the problem at hand. Under this novel perspective, it is no longer necessary to introduce modified trellises in order to perform decoding or to explicitly try to invert the syndrome formation procedure used during encoding, as done in \cite{liveris_DistributedCompressionOf} and in \cite{tu_SF-ISF} respectively. Differently from the other contributions on this subject, decoding is hence described in both parity- and syndrome-based approaches using the same factor-graph-based approach commonly taken in the LDPC-codes-related literature \cite{kschischang_FactorGraphsAnd}. Consequently, both binary (BSC and BEC) and non-binary (e.g.~\emph{additive white Gaussian noise}) transmission channels are handled under a unified formulation; moreover non-binary side information is handled as well. A similar result is reported in \cite{roumy_RateAdaptiveTurbo}, but employing \emph{ad hoc} encoding and decoding techniques on modified trellises.

The rest of the paper is organized as follows. In Section \ref{s:review} we review the practical approaches to the SCSI problem appeared so far, namely the parity- and the syndrome-based approaches, and investigate their theoretical performance when there exists a virtual BSC between the source and the side information. Section \ref{s:contribution} particularizes the parity- and the syndrome-based approaches to the case where turbo codes are employed, and shows how standard turbo encoding and decoding algorithms can be practically applied for data compression. In Section \ref{s:experiments} we show the compression performance of the discussed algorithms in a variety of settings, and compare them against other results in the literature referring to both parity- and syndrome-based systems. Concluding remarks on this work are given in Section \ref{s:conclusion}.

\section{Theoretical Limits and Connections to Channel Coding}\label{s:review}
Let a data source emit \emph{independent and identically distributed} (i.i.d.) realizations of a \emph{pair} of \emph{correlated} discrete-alphabet \emph{random variables} (r.v.) $(X,Y)$. From the source coding theorem \cite{shannon1948}, a sequence of $n$ of these realizations can be encoded (with an arbitrarily small probability of decoding error) using on average $R$ bit/realization iff $R\geq H(X,Y)$ and $n$ is sufficiently large, where $H(\cdot,\cdot)$ denotes the \emph{joint entropy} \cite{cover_ElemInfoThNew}.

Surprisingly, if the encoder was made by two independent components that cannot communicate with each other, one for encoding $X$ and the other for encoding $Y$, then the lower bound for the total rate would be the same. In particular, a sequence of $n$ joint realizations can be encoded (with an arbitrarily small probability of decoding error) using on average $R_X$ and $R_Y$ bit/sample by the $X$- and by the $Y$-component of the \emph{distributed} encoder, respectively, iff $R_X + R_Y \geq H(X,Y)$, $R_X \geq H(X|Y)$, $R_Y \geq H(Y|X)$, and $n$ is sufficiently large, where $H(\cdot|\cdot)$ denotes the \emph{conditional entropy} \cite{slepian_noiselessCodingCorrSrc}.

Consequently, in the problem of lossless SCSI, where $X$ is encoded with $Y$ being perfectly known at the decoder (i.e.~$R_Y \geq H(Y)$), the lower bound for the source coding rate $R_s$ is $R_s \geq H(X|Y) \leq H(X)$. In order to actually construct a coding system that reaches this limit, the lossless SCSI problem was first recast as a channel coding problem. The interpretation of $X$ and $Y$ as inputs or outputs of a \emph{virtual} \emph{correlation channel} (CC) is the key that led to this connection.

The vast majority of the literature discusses the binary case and assumes that $X$ and $Y$ are connected by a BSC. More precisely, the side information $Y$ is seen either as the \emph{output} (``forward'' BSC model) or as the \emph{input} (``backward'' BSC model) of the BSC channel. In the first case there exist a r.v.~$Z^f$ independent from $X$ such that $Y = X \oplus Z^f$; in the second one there exist a r.v.~$Z^b$ independent from $Y$ such that $X = Y \oplus Z^b$.\footnote{In both cases it is assumed that $p^\ast \triangleq P[Z^\ast = 1]$ satisfies $0 < p^\ast < 1/2$, which is not restrictive. In fact, if $p^\ast = 0$ then $X=Y$, if $p^\ast = 1/2$ then $X$ and $Y$ are actually independent, and if $p^\ast > 1/2$ then the CC could be simply seen as a BSC with error probability $1-p^\ast < 1/2$ followed by a \emph{deterministic} symbol inversion.} However, it is often also assumed that the source $X$ is \emph{uniformly distributed} (u.d.). In both cases, this implies (i) that $Y$ is u.d.~as well, and (ii) that it does not actually matter if the BSC is seen in one or in the other direction (i.e.~$Z^f$ or $Z^b$ turns out to be independent from both $X$ and $Y$).

In this paper, instead, we focus on the case where $X$ can possibly be non-u.d., i.e.~where the CC being a ``forward'' or ``backward'' BSC actually matters and leads to different considerations, from both the theoretical and the practical point of view. In the rest of this section, we will take the theoretical perspective, and discuss the implications of the two models with both the parity and the syndrome approaches.

\subsection{Forward BSC Model}\label{s:fw}
According to this model, (i) the lower bound for compressing $X$ is $H(X|Y) = H(Y|X) - [H(Y) - H(X)] = H(Z^f) - [H(Y) - H(X)]$, which satisfies $H(X|Y) < H(Z^f)$ (unless $X$ is u.d.), and (ii) the \emph{unconstrained} capacity of the virtual BSC is $C^f = 1 - H(Z^f) < 1$ bit/channel use. Since the capacity of a BSC can be approached by \emph{linear codes} \cite{mackay_inftheory}, and linear codes can always be generated by a \emph{systematic} encoder, the first devised strategy for the lossless SCSI problem was the \emph{parity-based approach}.

\subsubsection{Parity-based approach}\label{s:fw-parity}
First, a linear $(n,k)$ code approaching the capacity of the virtual BSC, i.e.~such that it achieves an arbitrarily small probability of decoding error with $R_c = k/n \simeq C^f$, is taken (in general, $R_c \leq C^f$). Then, each successive sequence of $k$ realizations from $X$ is fed into the corresponding channel encoder, that computes $n - k$ \emph{parity bits}. These bits form the compressed representation and are sent to the channel decoder. In turn, the channel decoder ``receives'' as well the corresponding $k$ realizations from $Y$ and reconstructs the ``transmitted'' codeword, made by the $k$ bits from the source (if no decoding errors have occurred) and by the $n - k$ parity bits (which are simply discarded).

The compression rate achieved by the parity-based approach is $R_s = \frac{n - k}{k} = \frac{n}{k} (1 - R_c)$ bit/sample\footnote{As a remark, note that even if $X$ is not u.d.~the parity message \emph{is i.i.d.~and u.d.}, at least asymptotically with the codelength $n$, so that no further rate reduction is possible as long as the parity message is sent losslessly.}. Consequently, $R_s \geq \frac{n}{k} H(Z^f) > H(Z^f) \geq H(X|Y)$, showing that this approach cannot achieve the theoretical bound, not even in the case of $X$ being u.d.

Nevertheless, in this scenario the channel decoder could actually ``correct'' losses in the parity message as well, which is more than what needed for losslessly decoding $X$. For example, as long as the parity message is received as if it was gone through a BSC with error-rate less or equal to the one of the virtual BSC, the decoder would still be able to reconstruct $X$ with arbitrarily small probability of error. Hence, either we conclude that the approach is suboptimal, but somewhat robust against error-prone transmission of the parity message, or we deliberately employ lossy compression of the parity message at the encoder\footnote{Since the parity message is i.i.d.~and u.d., the effect of optimal lossy compression can be taken into account by assuming the existence of a ``quantization'' BSC between the true parity message and the one sent to the decoder \cite{cover_ElemInfoThNew}.}, achieving an actual rate reduction.

The maximum rate reduction of this \emph{quantized} parity-based approach is achieved when the error-rate of the ``quantization'' BSC is equal to the one of the virtual BSC. In this case, the rate is reduced by a factor $1 - H(Z^f) = C^f$, leading to a compression rate $R_s^Q = C^f R_s \geq H(Z^f) \geq H(X|Y)$, where the first inequality is an equality iff the code achieves capacity and the second inequality is an equality iff $X$ is u.d. Finally, we conclude that the quantized parity-based approach achieves the theoretical bound, but only in the case of $X$ being u.d.\footnote{In this paper, both for the forward and for the backward case, we always assume that the channel decoder is informed about (i) the exact encoding process (which code, parity or syndrome, quantization used or not, \dots), (ii) the statistics of the \emph{transmission channel} (TC) between encoder and decoder, and (iii) the statistics of the virtual BSC, but not necessarily about the statistics of $X$. If this statistics was known at the decoder, then the actual channel code (over which the decoder conducts its search) would be a subset of the linear code used for parity formation, and our conclusions would be incorrect (namely, less parity bits could be sufficient for correct decoding).}

Alternatively, the parity message could be formed with respect to an higher-rate \emph{ad-hoc} code achieving the capacity of the \emph{true} channel, over which the parity is known to be not harmed. The average capacity of this channel is $C^f_t = \frac{k}{n} [1 - H(Z^f)] + \frac{n-k}{n} = 1 - \frac{k}{n} H(Z^f)$, so that the channel coding rate constraint could be relaxed to $R_c \leq C^f_t > C^f$, that in turn would lead to $R_s \geq H(Z^f) \geq H(X|Y)$, i.e.~to the same conclusions obtained for the quantized parity-based approach. However, it turns out that this approach is rather a syndrome-based and not a parity-based one. In fact, the discussion about the syndrome-based approach in the following will show that the parity message with respect to an ad-hoc code could be simply seen as a syndrome message.

\subsubsection{Syndrome-based approach}
Again, take a linear $(n,k)$ code approaching the capacity of the virtual BSC. This code partitions the set of all sequences of $n$ symbols from the input alphabet into $2 ^ {n - k}$ \emph{cosets} that are as good as the original code for channel coding. Then, in correspondence of each successive sequence of $n$ realizations from $X$, the encoder identifies the coset to which that sequence belongs. This information is encoded as $n - k$ \emph{syndrome bits} that form the compressed representation and are sent to the decoder. Upon decoding (in principle) the corresponding $n$ realizations from $Y$ (i.e.~a \emph{corrupted} codeword) into the signalled coset, the $n$ bits from the source can be eventually reconstructed with an arbitrarily small probability of error.

As noted in \cite{tan_SWcodingParitySyndrome}, from an $(n,k)$ linear code used in a syndrome-based SCSI system an ad-hoc $(2n - k,n)$ systematic encoder can be derived for an equivalent parity-based SCSI system. In fact, the $n - k$ bits used to specify the coset information can be seen as parity bits.

\begin{figure}
\centering
\includegraphics[scale=.5]{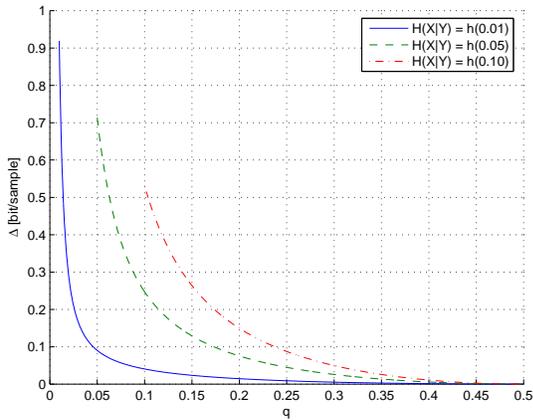}
\caption{Compression rate loss under the forward BSC model when $X$ is not u.d. ($h(\cdot)$ denotes the binary entropy function).}
\label{f:rateloss}
\end{figure}

Differently from the parity-based approach, in the syndrome-based approach it is not necessary to employ quantization in order to achieve the compression limit. In fact, the compressed representation requires now $R_s = \frac{n - k}{n} = 1 - R_c$ bit/sample\footnote{Similarly to the parity message, and at least asymptotically with the codelength $n$, the syndrome message \emph{is i.i.d.~and u.d.} even if $X$ is not u.d.}. Hence, $R_s \geq H(Z^f)$ (with equality iff the code achieves capacity), and in turn $H(Z^f)$ equals $H(X|Y)$ iff $X$ is u.d. If we let $q$ denote the probability of $X$ being one, the compression rate loss $\Delta = H(Z^f) - H(X|Y)$ in correspondence of a fixed value of $H(X|Y)$ is shown in Fig.~\ref{f:rateloss}. We conclude that the syndrome-based approach achieves the theoretical bound, but again only in the case of $X$ being u.d.

For example, the $(3,1)$ Hamming code (i.e.~the $(3,1)$ repetition code) is seen as a linear code achieving the capacity of the additive channel (on $GF(2^3)$) in which $Z^f$ is such that $p^f(000) = p^f(001) = p^f(010) = p^f(100) = 1/4$ \cite{pradhan_DISCUS1,xiong_DSCSensors}; in fact, this code can correct these error patterns and $C^f = 3 - H(Z^f) = 1$ bit/channel use. The $4$ cosets of the partition are $\{000,111\}$, $\{001,110\}$, $\{010,101\}$, and $\{011,100\}$. While it is usually stated that this syndrome-based system achieves $H(X|Y)$, it should be emphasized that under a forward BSC model this is true iff $X$ is u.d. As example, if $p_X(000) = p_X(001) = p_X(010) = p_X(011) = 1/16$, and $p_X(111) = p_X(110) = p_X(101) = p_X(100) = 3/16$, the information about the coset sent by the encoder requires $R_s = H(Z^f) = 2$ bit/sample (and no less than this), but $p_Y(000) = p_Y(001) = p_Y(010) = p_Y(011) = 3/32 = 1.5/16$ and $p_Y(111) = p_Y(110) = p_Y(101) = p_Y(100) = 5/32 = 2.5/16$, i.e.~$Y$ has a strictly greater entropy than $X$. Hence, $R_s = H(Z^f) > H(X|Y)$. \footnote{But, if the statistics of $X$ was known at the decoder, then the actual code (over which the decoder conducts its search) would be a subset of the coset specified by the syndrome message. Our conclusions would be in this case incorrect (namely, cosets could be made ``larger'' and still achieve correct decoding, while requiring less bits for the corresponding syndrome message).}

\subsection{Backward BSC Model}
According to this model, the lower bound for compressing $X$ is simply $H(X|Y) = H(Z^b)$; the \emph{unconstrained} capacity of the virtual BSC is $C^b = 1 - H(Z^b) < 1$ bit/channel use. Despite the different correlation structure, which in case of $X$ being not u.d.~is \emph{really} different, both the parity- and the syndrome-based encoding procedures (with respect to a linear code achieving the capacity of the virtual BSC) can be employed exactly as in the case of the forward BSC model. However, the decoding algorithm can be now related to the channel decoding strategy that would be used for reliable transmission over the virtual BSC only in the case of the syndrome-based approach.

In particular, since the syndrome in correspondence of each successive sequence of $n$ realizations from $X$ can be generated by means of a \emph{linear} function, and the decoder can similarly generate the syndrome of the corresponding sequences of $n$ realizations from $Y$, the decoder can first compute the syndrome of the difference. Hence, it actually knows the coset into which the difference $Z^b$ lies. But the code and its cosets are such that the \emph{typical} errors across the virtual BSC can be corrected. Eventually, the actual difference (and the actual source) can be found with arbitrarily small probability of error as the only typical element in that particular coset.

The compression rate achieved by the syndrome-based approach is again $R_s = \frac{n - k}{n} = 1 - R_c$ bit/sample. But, differently from the forward BSC case, $R_s \geq H(Z^b) = H(X|Y)$, with equality iff the code achieves capacity. Hence, we conclude in addition that the syndrome-based approach achieves the theoretical SCSI bound, independently from the actual statistics of $X$.

Despite in the case of the parity-based approach the connection to the channel decoding strategy for the virtual BSC is less straightforward, this does not represent a problem. In fact, the optimal SCSI decoder, as discussed in the next section, can be readily derived as solution of a \emph{maximum a posteriori probability} (MAP) problem. In addition, it turns out that the decoder can still re-use the typical channel decoding algorithms that would be used for reliable transmission over the virtual BSC, confirming once again that the SCSI problem is in fact a channel coding problem.

\subsubsection{Compression without Side Information}
The SCSI problem is obviously an extension of the simpler source coding problem (without side information). In particular, the source coding problem falls exactly into the backward BSC model. In fact, it is sufficient to assume that a ``fake'' side information $Y$ exists, which is identically zero, and that $Z^b = X$. Hence, all conclusions about the SCSI algorithms applied to the backward BSC model hold for the simpler problem of source compression too.

\section{SCSI and Data Compression Using Turbo Codes}\label{s:contribution}
From the discussion above, it appears that the search for ``good'' SCSI systems reduces to the search for ``good'' channel codes. As \emph{turbo} codes \cite{berrou_turboJCOM,mackay_inftheory} come very close to achieving the promise of Shannon's channel capacity theorem, many SCSI systems appeared in literature take advantage of their application.

\subsection{Turbo-Parity Formation}
The conventional (parallel concatenated) \emph{turbo encoder} is a \emph{systematic} encoder: in correspondence of a sequence of $Nk$ realizations from $X$ ($\bm{x}$) it uses two systematic $(n,k)$ \emph{convolutional} codes to form two sequences of parity bits of $N(n - k) + z_t$ bits each\footnote{The additional $z_t \ll N(n - k)$ bits are emitted for terminating the encoding into the zero state \cite{richardson_mct}.} ($\bm{p}_0$ and $\bm{p}_1$). The source bits enter the second convolutional encoder, that can also be equal to the first,  after being randomly interleaved (reordered).

The turbo code can hence be seen as a giant $(N(2n - k) + 2z_t,Nk)$ systematic \emph{block} code whose \emph{generator matrix} is
\begin{equation}
\bm{G} = \left[\begin{array}{c|c|c}
\bm{I}_{Nk} & \bm{P}_0 & \bm{P}_1 \\
\end{array}\right]\;,\nonumber
\end{equation}
where $\bm{P}_i$ is the $Nk \times [N(n - k) + z_t]$ \emph{parity formation} matrix corresponding to the $i$-th convolutional code (comprehensive of the interleaver if $i = 1$).

Before sending the parity to the decoder, \emph{puncturing} (i.e.~bit removal) can be employed for rate adaptation at the encoder. The encoder can hence operate at any rate $0 \leq R_s \leq \frac{2(n - k)}{k}$; in particular, for appropriate choices of $(n,k)$, rates greater than $1$ bit/sample are possible too. If puncturing is employed, then the equivalent generator matrix is $\bm{G}' = [\bm{I}_{Nk} | \bm{P}_0' | \bm{P}_1']$, where $\bm{P}'_i$ is the $Nk \times s_i$ matrix obtained removing from $\bm{P}_i$ the columns corresponding to the punctured parity bits.

\subsection{Turbo-Syndrome Formation}\label{s:syn_generation}
From the generator matrix of the turbo code, the parity-check matrix
\begin{equation}
\bm{H}' = \left[\begin{array}{c|c|c}
\bm{P}_0'^T & \bm{I}_{s_0}            & \bm{0}_{s_0 \times s_1} \\
\hline
\bm{P}_1'^T & \bm{0}_{s_1 \times s_0} & \bm{I}_{s_1}            \\
\end{array}\right]\nonumber
\end{equation}
is immediately derived that can be used for syndrome formation. All other parity check matrices can be derived from $\bm{H}'$ by left-multiplication with an invertible $(s_0 + s_1) \times (s_0 + s_1)$ matrix: in case of an error-prone TC it is possible that some of them lead to a better performance than $\bm{H}'$ \cite{tan_EnhancingTheRobustness,roumy_RateAdaptiveTurbo}.

However, if $\bm{H}'$ is employed, in correspondence of a sequence of $Nk + s_0 + s_1$ outcomes from $X$ (partitioned into the three sub-sequences $\bm{x}$, $\bm{x}_0$, and $\bm{x}_1$ of length $Nk$, $s_0$, and $s_1$ respectively), the syndrome (i.e.~the right-multiplication of $[\bm{x} | \bm{x}_0 | \bm{x}_1]$ by $\bm{H}'^T$) can be simply obtained by (i) forming the (punctured) parities $\bm{p}_0$ and $\bm{p}_1$ corresponding to $\bm{x}_0$, and (ii) adding $\bm{p}_i$ with $\bm{x}_i$. The syndrome message $[\bm{s}_0, \bm{s}_1]$ is hence made of two sequences $\bm{s}_i = \bm{p}_i \oplus \bm{x}_i$ of $s_i$ bits each that are directly obtained re-utilizing the conventional turbo encoder.

The encoder can operate at any rate $0 \leq R_s \leq \frac{2(n - k)}{2n - k} < 1$ bit/sample; in particular, no rates greater than $1$ bit/sample are realizable: for example, if $(2,1)$ constituent codes are employed, then $0 \leq R_s \leq 2/3$ bit/sample.

\subsection{Unified Decoding}
From the discussion in Section \ref{s:review}, it may seem that the actual SCSI decoding procedures cannot be exactly the ones employed in channel decoding, and also must depend from the correlation model and from the particular message received at the decoder (parity or syndrome). In fact, many contributions on this subject proposed ad-hoc decoding techniques, tailored for the specific settings treated. However, as shown in \cite{cappellari10_onSyndromeDecodingCOML}, these techniques are actually in most cases the same and could be simply derived by tackling the problem as a MAP one. In the turbo case, this strategy immediately indicates how to re-utilize the conventional turbo decoder in the SCSI decoder.

In the parity case, assuming that the parity messages $\bm{p}_i$ are received at the decoder as $\bm{r}_i$, and that $\bm{y}$ denotes the $Nk$-dimensional realization of $Y$ corresponding to the source realization $\bm{x}$ being compressed, the optimum MAP estimate is simply found as\footnote{Given two r.v.~$A$ and $B$, the \emph{likelihood} and \emph{a posteriori probability} (APP) functions will hereafter be denoted by $l_a(b) \triangleq p(a|b)$ and $p_a(b) \triangleq p(b|a)$ respectively, so that the ``free'' variable always appears as argument of the function while the known one always appears as subscript (parameter).}
\begin{equation}
\arg \max_{\bm{x}} p\left(\bm{x}|\bm{y}\bm{r}_0\bm{r}_1\right) = \arg \max_{\bm{x}} p_{\bm{y}\bm{r}_0\bm{r}_1}(\bm{x})\;.\nonumber
\end{equation}
This probability can be obtained marginalizing the function $p_{\bm{y}\bm{r}_0\bm{r}_1}(\bm{x}\bm{p}_0\bm{p}_1)$ that, apart some scaling factors, factorizes into the product of (i) $p(\bm{x}\bm{y})$, (ii) $\chi_i(\bm{p}_i|\bm{x})$, that are indicator functions unitary iff $\bm{p}_i$ is the actual parity of $\bm{x}$ (according to the $i$-th convolutional code, and comprehensive of the interleaver if $i = 1$), and (iii) $l_{\bm{r}_i}(\bm{p}_i)$, that take into account for the TC. As a remark, note that if $\bm{y}$ is seen as the systematic portion of the codeword ``received'' by the decoder, this maximization is exactly the one performed by an optimal MAP channel decoder, i.e.~there is no difference between a turbo decoder and a SCSI decoder.

\begin{figure}
\centering
\subfigure[parity-based approach]{
\includegraphics[scale=.85]{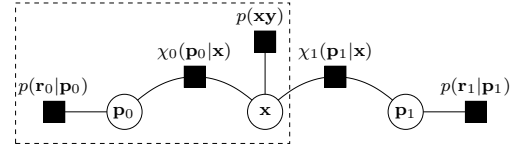}
\label{f:graph_parity}
}\\
\subfigure[syndrome-based approach]{
\includegraphics[scale=.85]{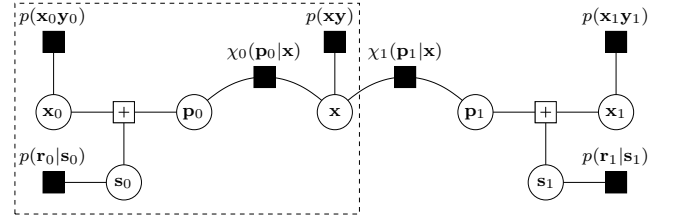}
\label{f:graph_syndrome}
}
\caption{Factor graphs for iterative decoding in the problem of SCSI using turbo codes. At each turbo iteration involving one of the two constituent codes (e.g.~the one whose function to be marginalized is represented by the sub-graph in the dashed box), the incoming message across the box is seen as prior information about $\bm{x}$.}
\label{f:factor-graph}
\end{figure}

In practice, for turbo codes, an iterative procedure is used for approximating MAP decoding that is easily described as a \emph{message-passing} algorithm on the \emph{factor graph} representing this factorization, shown in Fig.~\ref{f:graph_parity} (for a useful tutorial article on factor graphs and message-passing algorithms, the reader is referred to \cite{kschischang_FactorGraphsAnd}). The traditional decoding algorithm employs the \emph{forward-backward algorithm} (also known as BCJR algorithm \cite{bahl_OptimalDecodingOf}) in order to exactly marginalize the function relative to one constituent code (represented by the sub-graph in the dashed box, which has no cycles) using the message incoming from the previous iteration (involving the other constituent code) as additional \emph{prior} information about $\bm{x}$, and produces a new prior for the next iteration.

In the syndrome case, assuming that the syndrome messages $\bm{s}_i$ are received at the decoder as $\bm{r}_i$, and that the sequence of $Nk + s_0 + s_1$ realizations from $Y$ corresponding to the source realization being compressed is equivalently partitioned into the three sub-sequences $\bm{y}$, $\bm{y}_0$, and $\bm{y}_1$, the optimum MAP estimate is found as
\begin{equation}
\arg \max_{\bm{x}\bm{x}_0\bm{x}_1} p_{\bm{y}\bm{y}_0\bm{y}_1\bm{r}_0\bm{r}_1}(\bm{x}\bm{x}_0\bm{x}_1)\;.\nonumber
\end{equation}
This probability can be now obtained marginalizing the function $p_{\bm{y}\bm{y}_0\bm{y}_1\bm{r}_0\bm{r}_1}(\bm{x}\bm{x}_0\bm{x}_1\bm{p}_0\bm{p}_1\bm{s}_0\bm{s}_1)$ that, apart some scaling factors, factorizes into the product of (i) $p(\bm{x}\bm{y})p(\bm{x}_0\bm{y}_0)p(\bm{x}_1\bm{y}_1)$, (ii) $\chi_i(\bm{p}_i|\bm{x})$, (iii) $\chi_{\{\bm{p}_i \oplus \bm{x}_i = \bm{s}_i\}}$, that are indicator functions of the condition in brackets, and (iv) $l_{\bm{r}_i}(\bm{s}_i)$. The only difference w.r.t.~the parity case is represented by the introduction of the factors in (iii). Most importantly, these factors do not add any cycle in the factor graph, as shown in Fig.~\ref{f:graph_syndrome}. Hence, decoding can be performed re-using the turbo decoding algorithm presented above. In particular, it is only necessary to form the correct input likelihoods to the parity nodes $\bm{p}_i$ and then post-process their final APP approximation for obtaining the APP approximation for the source nodes $\bm{x}_i$. In the following, this decoding procedure is referred as \emph{soft-syndrome decoding} (SSD), while the suboptimal SCSI decoding procedure resulting from employing a turbo decoder that communicates only the \emph{hard} choices for $\bm{p}_i$ (and not their full APP functions) is referred as \emph{hard-syndrome decoding} (HSD).

If the CC is a forward BSC, $p(\bm{x}\bm{y}) = p(\bm{x})l_{\bm{y}}(\bm{x}) = p(\bm{x})p_{Z^f}(\bm{y} - \bm{x})$; if the CC is a backward BSC, $p(\bm{x}\bm{y}) \propto p_{\bm{y}}(\bm{x}) = p_{Z^b}(\bm{x} - \bm{y})$ (the same holds for $p(\bm{x}_i\bm{y}_i)$). In the first case, MAP decoding is feasible only if $p(\bm{x})$ is also known at the decoder. Since we assume that the decoder is not aware of this information, we actually operate it without that factor (i.e.~as if $X$ was u.d.), performing what is known as \emph{maximum likelihood} (ML) decoding. In both cases, the SCSI decoder re-utilizes in the best possible way the traditional turbo decoding algorithm, without needs for designing any particular parity/syndrome manipulation or inversion.

\subsection{Discussion}\label{s:discussion}
The turbo-syndrome formation algorithm described in Section \ref{s:syn_generation} corresponds to the algorithm used in \cite{zamani09_AFlexibleRate} and to the ``zero-forcing'' algorithm described in \cite{tu_SF-ISF} and in other papers by the same authors. While we directly tackle MAP decoding of the turbo-syndrome, the front-end of the decoder used in \cite{tu_SF-ISF} consists of a hard-in hard-out \emph{inverse syndrome former} (ISF). With binary TC-output and side information, the straight ISF-based decoder implements exactly what we named hard-syndrome decoding.

The factor-graph approach has certain advantages with respect to the utilization of an ISF. First, neither the TC-output nor the side information (in SCSI problems) are restricted to be binary in order to perform decoding. Then, in case of an error-prone TC (a BSC is for example tested in \cite{tan_EnhancingTheRobustness}), the optimum input likelihoods to the traditional turbo-decoding algorithm are immediately known without the need to analyze the signal flow through the ISF. Finally, syndromes which are not formed for both constituent codes according to the ``zero-forcing'' approach, for which the ISF is difficult, if not unfeasible, to construct, can be handled too. 

However, the ``zero-forced'' syndrome is not really robust against TC errors. In fact, while source bits belonging to $\bm{x}$ are effectively ``protected'' by the turbo-code, any source bit belonging to $\bm{x}_i$ participates only to a single check for syndrome formation, such that erasures or flips to a syndrome bit make the correct recovery of the corresponding source bit very hard. By using a polynomial parity-check matrix, as proposed in \cite{tan_EnhancingTheRobustness} and \cite{roumy_RateAdaptiveTurbo} for convolutional and turbo codes respectively, more robust syndromes have been found that in particular support puncturing. But, not only the resulting encoder can no longer rely on traditional turbo-encoding algorithms, but also efficient decoding must be actually performed on a more complex trellis (named \emph{super trellis} in \cite{roumy_RateAdaptiveTurbo}) that no longer shares the same transitions of the original trellis.

\section{Experiments and Comparisons}\label{s:experiments}
Experiments have been done under both the backward and the forward BSC correlation models. In the latter case, we focused on the case of non-u.d. sources. We also compared our results with many others from the literature.

\subsection{Experimental Setup}
The same turbo code and the same data frame length $L$ have been employed for both parity- and syndrome-based approaches. In particular, the turbo code uses two identical $(n,k)=(2,1)$, $16$-state, systematic constituent encoders with generator matrix $\bm{G}(D) = \left[1\ \frac{D^4+D^2+D+1}{D^4+D^3+1}\right]$, and a random interleaver in between. Two different frame lengths have been considered, namely $L=2^{14}=16384$ samples (``short'' frame) and $L=2^{16}=65536$ samples (``long'' frame). Random puncturing of the parity bits is employed for rate adaptation.

All decoding routines are set for a (maximum) number of runs of the FBA algorithm equal to $40$ ($20$ iterations for each code). However, in order to reduce the decoding complexity, a stopping criterion breaks the decoding task whenever both the constituent FBA algorithms indicate persistent and mutually consistent decoded codewords. As suggested in \cite{mackay_inftheory}, during each FBA run the most probable transition at each time-step is evaluated in order to check if the sequence of all such transitions forms a valid codeword. In practice, we consider the last $4$ consecutive FBA runs and check if in all of them the same codeword is obtained. If this condition is met, the turbo loop for the current frame is stopped and the last computed likelihoods are emitted.

Only error-free transmission channels had been considered in the simulations. For all choices of the simulation parameters, $2^{15}=32768$ or $2^{13}=8192$ frames (in the short and in the long case, respectively) have been generated, such that the average \emph{bit error ratio} (BER) is eventually estimated over $2^{29}\simeq 5\cdot 10^8$ samples.

\subsection{Backward BSC}
\begin{figure}
\centering
\includegraphics[scale=.5]{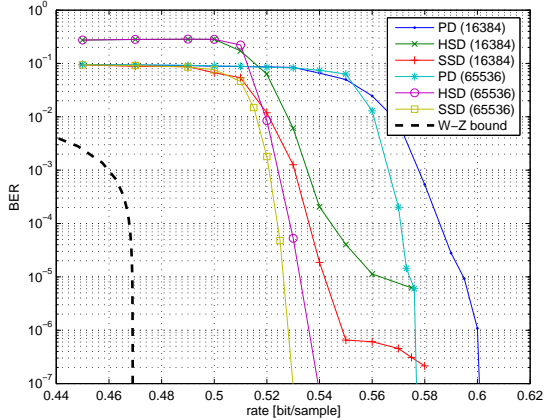}
\caption{Comparison between the different SCSI algorithms based on turbo decoding, for a backward BSC model with $p^b = 0.10$ (and \emph{any} source statistics).}
\label{f:data_xx_10}
\end{figure}

\begin{figure}
\centering
\includegraphics[scale=.5]{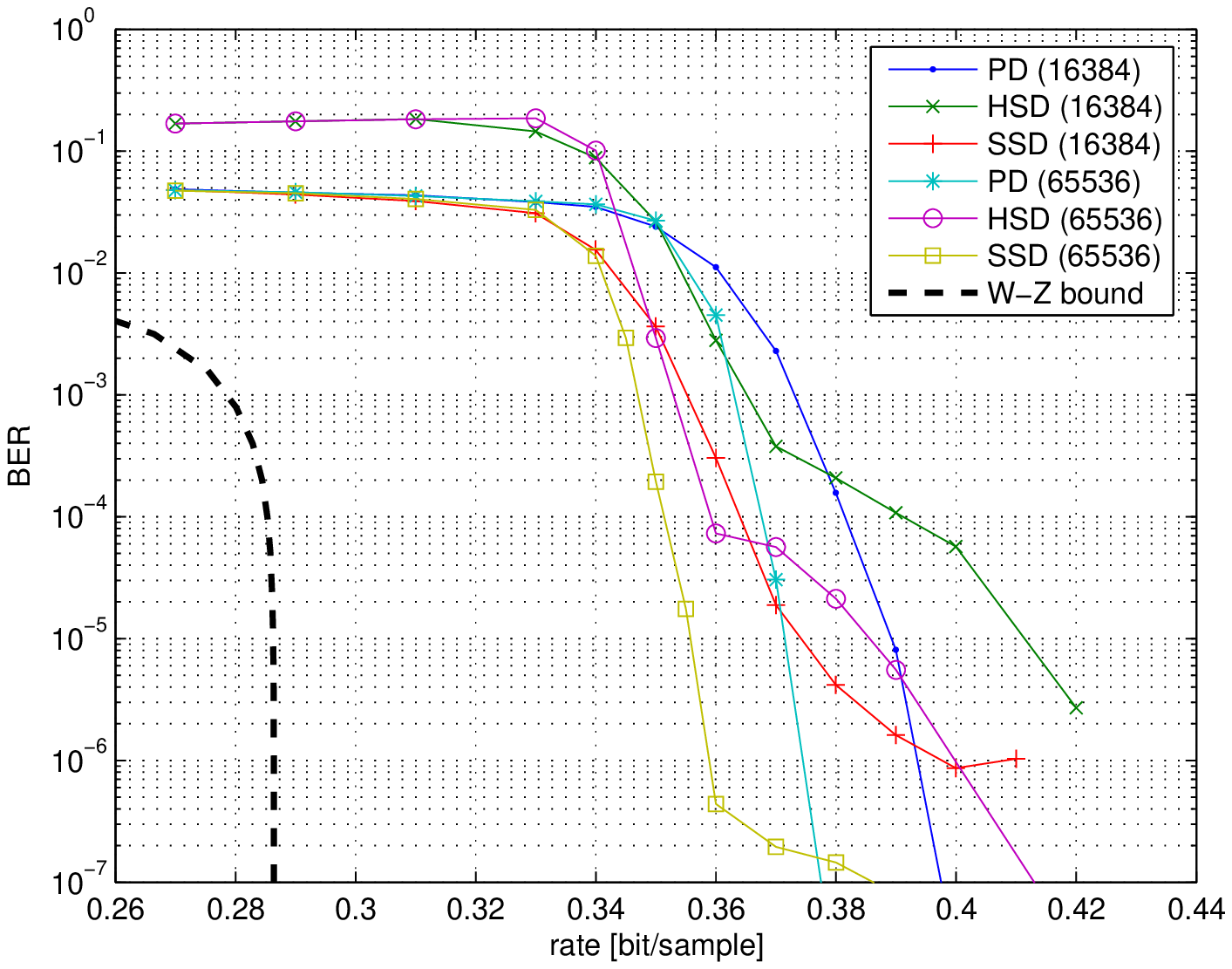}
\caption{Comparison between the different SCSI algorithms based on turbo decoding, for a backward BSC model with $p^b = 0.05$ (and \emph{any} source statistics).}
\label{f:data_xx_05}
\end{figure}

\begin{figure}
\centering
\includegraphics[scale=.5]{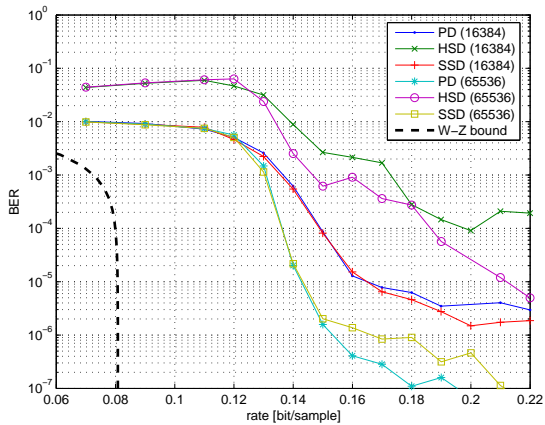}
\caption{Comparison between the different SCSI algorithms based on turbo decoding, for a backward BSC model with $p^b = 0.01$ (and \emph{any} source statistics).}
\label{f:data_xx_01}
\end{figure}

We fixed the value of $p^b$ and measured the performance of the considered decoding algorithms as a function of the target coding rate $R_s$. Three cases were considered, namely $p^b = 0.10$, $p^b = 0.05$, and $p^b = 0.01$. The simulation results are reported in Fig.~\ref{f:data_xx_10}, Fig.~\ref{f:data_xx_05}, and Fig.~\ref{f:data_xx_01}, respectively.

In all cases, the compression limit is given by $H(Z^b)=h(p^b)$. This limit is independent from the source statistics. In fact, despite the curves shown refer to a uniform distribution, we also checked that the same exact results are obtained with \emph{any} distribution. In the plots we show the theoretical limit in terms of the rate-distortion function (``Wyner-Ziv bound'') \cite{wyner_RDFunctSrcCodingSideInfo}\footnote{Despite this function is derived for uniform source and side distribution, we use it also in the non-uniform cases. Note that, however, for BER$\to 0$ this function converges to $H(X|Y)$, which is the lossless compression limit independently from the source statistics.}. The curve labelled ``PD'' refers to parity decoding, while the other ones refer to syndrome decoding, \emph{hard} or \emph{soft}. The length of the frames is given in parentheses.

The SSD approach always presents a waterfall region closer to the Wyner-Ziv bound than the PD approach of the same length does. The gap between these curves tends to disappear when source and side information are very correlated ($p^b\to 0$). This seems to suggest that the parity-approach may have in this case a theoretical limitation similar to the one found in Section \ref{s:fw-parity} for the forward BSC case. The factor $n/k>1$ responsible for the gap would in fact be closer to one (i.e.~no loss) when less parity bits are formed.

Despite the waterfall region of SSD is closer to the Wyner-Ziv bound, the \emph{error floor} associated to SSD is also higher than the one associated to PD, especially for the high correlation case. As the puncturing increases (i.e.~as $R_s$ decreases) both PD and SSD present higher error floor regions and more irregular BER curves, probably due to the \emph{heavy} and \emph{unoptimized} puncturing of the parity of both constituent codes. This behavior is much more visible in the syndrome-based approach than it is in the parity-based approach, as can be seen, in particular, in Fig.~\ref{f:data_xx_01}.

In both approaches, a sharper waterfall curve and a better performance are obtained with long frame lengths rather than with short frame lengths. This fact is reasonable since large interleaver lengths are likely to generate more randomly distributed codewords.

The HSD approach has a performance in between the one of SSD and the one of PD, at least for low correlations, but its error floor is rather high already for $p^b=0.05$. Instead, for high correlation (for example when $p^b=0.01$), it is worth to note that HSD performs very poorly with respect to both SSD and PD.

\subsubsection{Comparison with Other Systems}
The results obtained for the backward BSC correlation model hold for any source distribution. They can hence be compared also with results that refer to a forward BSC model, at least as long as a uniform source distribution is assumed in the latter case.

In Table \ref{t:compression_rates}, this comparison is given in terms of rate required for near-lossless compression. For systems based on channel codes, where a residual error is always expected, a BER $\leq 10^{-6}$ is considered as the threshold for almost perfect reconstruction. The rates reported in the Table consider the case $p^b = 0.10$ or $p^b = 0.05$, and are divided in two sections, the first for parity-based methods and the second for syndrome-based ones. In both sections, the methods are sorted according to their average performance under the two working conditions.

\begin{table}
\renewcommand{\arraystretch}{1.3}
\caption{Comparison between different SCSI methods: the minimum compression ratio such that BER $\leq 10^{-6}$ is showed, for $p^b = .10$ and $p^b = .05$. In parentheses, the gap from the theoretical limit is showed. The frame length is reported too. In both parts of the Table (referring to parity- and syndrome-based approaches respectively) the various methods are sorted by increasing performance.}
\label{t:compression_rates}
\centering
\begin{tabular}{c||c|c}
\hline
$p^b$    & $.10$ & $.05$ \\
$H(Z^b) = h(p^b)$ & $.469$ & $.286$ \\
\hline
\hline
bzip2 ($16384$) \cite{garcia_CompressionOfBinary}                    & $.670\ (+.201)$ & $.440\ (+.154)$ \\
Turbo parity, $8$-state ($16384$) \cite{garcia_compressionTurbo}    & $.630\ (+.161)$ & $.435\ (+.149)$ \\
Turbo parity ($10000$) \cite{Hagenauer2004}                          & $.590\ (+.121)$ & $.440\ (+.154)$ \\
gzip ($16384$) \cite{garcia_CompressionOfBinary}                     & $.600\ (+.131)$ & $.410\ (+.124)$ \\
PD, $16$-state ($16384$)                               & $.600\ (+.131)$ & $.394\ (+.108)$ \\
Turbo parity, $8$-state ($16384$) \cite{garcia_CompressionOfBinary} & $.580\ (+.111)$ & $.380\ (+.094)$ \\
PD, $16$-state ($65536$)                               & $\bm{.576\ (+.107)}$ & $\bm{.374\ (+.088)}$ \\
\hline
\hline
LDPC ($16384$) \cite{liveris_sideLDPC}                               & $.600\ (+.131)$ & $.402\ (+.116)$ \\
SSD, $16$-state ($16384$)                              & $.549\ (+.080)$ & $.398\ (+.112)$ \\
P\&C trellis, $8$-state ($16384$) \cite{liveris_DistributedCompressionOf} & $.556\ (+.087)$ & $.388\ (+.102)$ \\
SSD, $16$-state ($65536$)                              & $\bm{.528\ (+.059)}$ & $\bm{.359\ (+.073)}$ \\
\hline
\end{tabular}
\end{table}

As brief comment to these results, we highlight the fact that in the first section of the Table (parity-based approaches) the ``short'' PD method performs only slightly worse than the ``Turbo parity'' method of the same frame length proposed in \cite{garcia_CompressionOfBinary}, which in turn is outperformed only by the ``long'' PD method. For what concerns the syndrome-based approaches, the method ``P\&C trellis'' \cite{liveris_DistributedCompressionOf} is placed between the ``short'' and the ``long'' SSD methods. Even though these comparisons can be considered a little unfair since systems are based on different convolutional codes with different number of states, and on frames of different sizes, these results have been reported in order to give an idea on how the considered decoding techniques behave with respect to other systems known in literature.

\begin{figure}
\centering
\includegraphics[scale=.5]{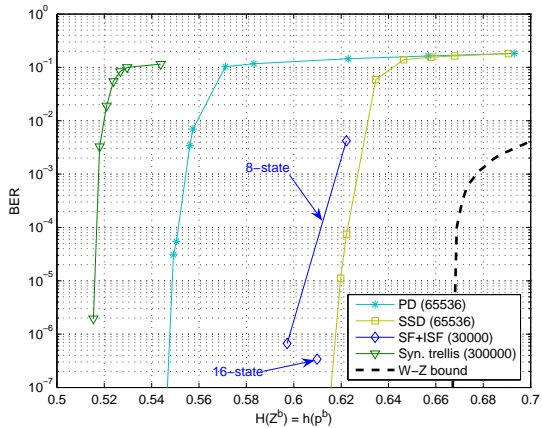}
\caption{Comparison between different SCSI methods, at fixed rate $R_s = 2/3$ bit/sample. The label ``SF+ISF'' refers to the syndrome-based method in \cite{tu_SF-ISF} (results for two different convolutional codes are shown); the label ``Syn. trellis'' refers to the syndrome-based method in \cite{roumy_RateAdaptiveTurbo}, where $16$-state constituent codes are employed. The frame length is reported in parentheses.}
\label{f:cmp_xx_r66}
\end{figure}

A more fair comparison is indeed given in Fig.~\ref{f:cmp_xx_r66}, in which the BER as a function of $h(p^b)$ is shown at rate $R_s = 2/3$ bit/sample. In this Figure, it can be seen that the proposed ``long'' SSD method outperforms the coding performance of the ``SF+ISF'' method given in \cite{tu_SF-ISF}. In fact, as observed in Section \ref{s:discussion}, these two methods are very similar but the latter is rather based on the sub-optimal HSD algorithm. Despite the very large interleaver length, the ``Syn.~trellis'' method proposed in \cite{roumy_RateAdaptiveTurbo} has instead a very poor performance, which is even worse than the performance of the ``long'' PD method.

\begin{figure}
\centering
\includegraphics[scale=.5]{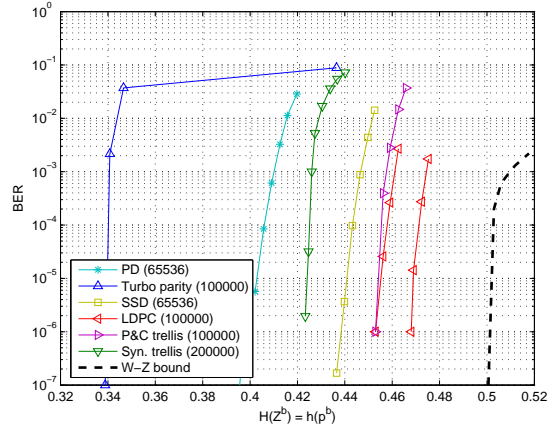}
\caption{Comparison between different SCSI methods, at fixed rate $R_s = 1/2$ bit/sample. The label ``Turbo parity'' refers to the parity-based method in \cite{aaron_compressionSideTurboCodes}, that uses two $(5,4)$ $16$-state constituent codes. The label ``LDPC'' refers to the syndrome-based method in \cite{liveris_sideLDPC} (results relative to two irregular LDPC codes are shown); the label ``P\&C trellis'' refers to the syndrome-based method in \cite{liveris_DistributedCompressionOf} that uses $16$-state constituent codes; the label ``Syn. trellis'' refers to the syndrome-based method in \cite{roumy_RateAdaptiveTurbo} ($16$-state). The frame length is reported in parentheses.}
\label{f:cmp_xx_r50}
\end{figure}

Finally, Fig.~\ref{f:cmp_xx_r50} shows some results relative to $R_s = 1/2$ bit/sample. In this case the proposed ``long'' SSD method has again a good performance, which is overcome only by the LDPC-based systems reported in \cite{liveris_sideLDPC} (which employ a longer frame size) and by the ``P\&C trellis'' method proposed in \cite{liveris_DistributedCompressionOf}, which makes use of longer frames and of different $16$-state constituent codes (specifically tailored for heavy data puncturing). Again, despite its error-resilience properties and the very long frame size, the ``Syn.~trellis'' method \cite{roumy_RateAdaptiveTurbo} has very poor performance.

\subsection{Forward BSC model}
\begin{figure}
\centering
\includegraphics[scale=.5]{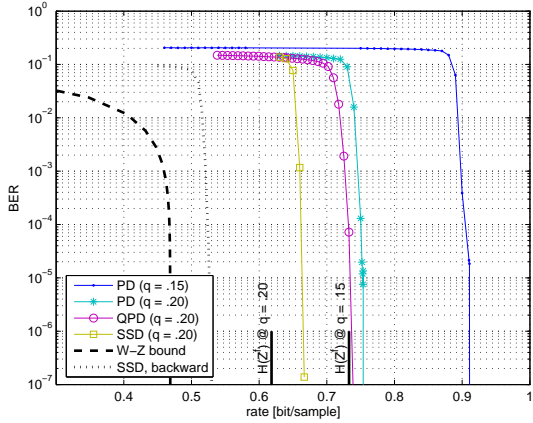}
\caption{Comparison between the different SCSI algorithms based on turbo decoding, for a forward BSC model with fixed $H(X|Y) = h(0.10)$, and $q = 0.15$ or $q = 0.20$.}
\label{f:data_FC_xx_10}
\end{figure}

\begin{figure}
\centering
\includegraphics[scale=.5]{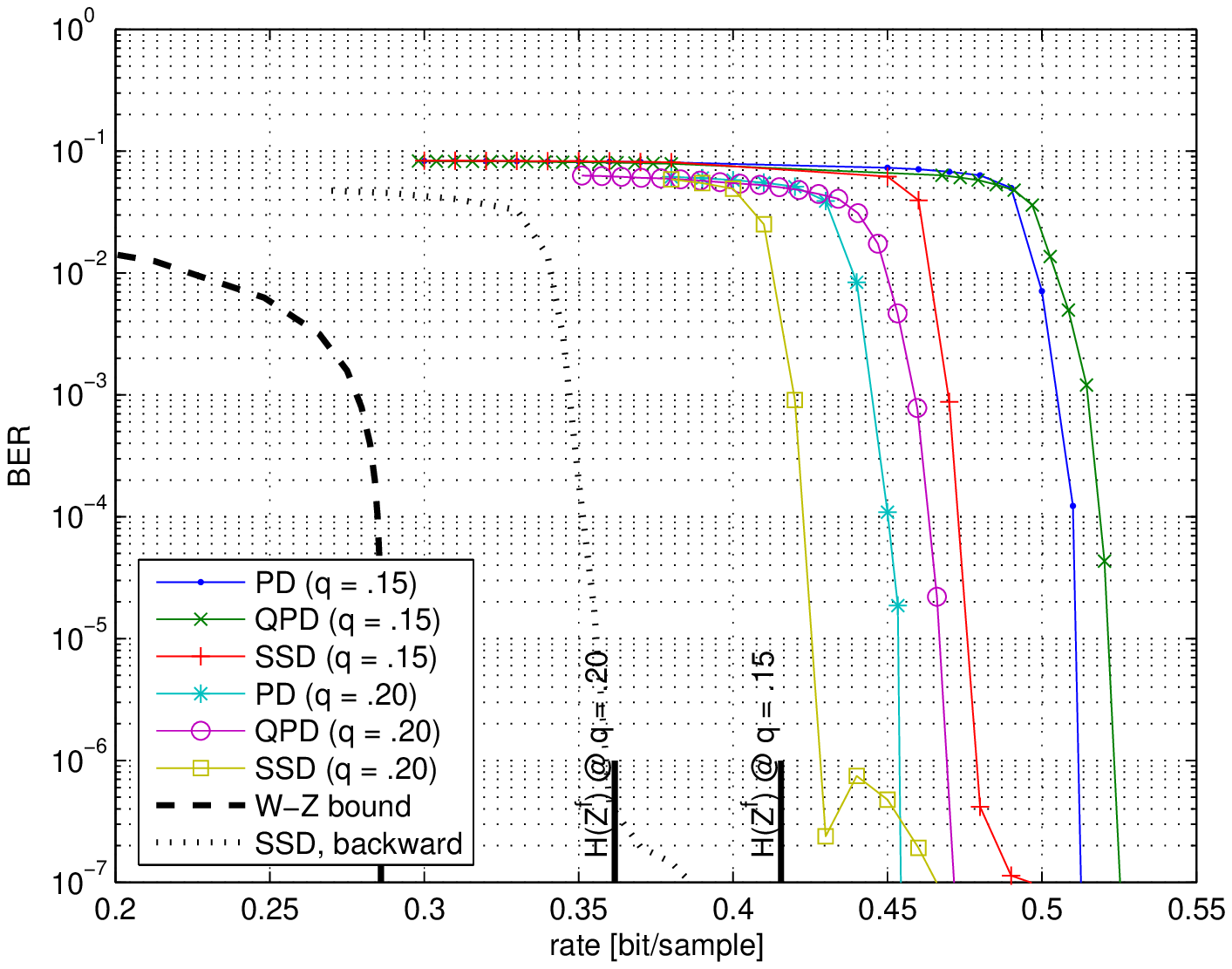}
\caption{Comparison between the different SCSI algorithms based on turbo decoding, for a forward BSC model with fixed $H(X|Y) = h(0.05)$, and $q = 0.15$ or $q = 0.20$.}
\label{f:data_FC_xx_05}
\end{figure}

\begin{figure}
\centering
\includegraphics[scale=.5]{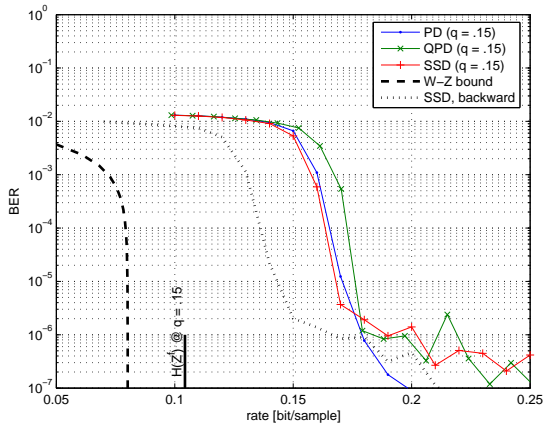}
\caption{Comparison between the different SCSI algorithms based on turbo decoding, for a forward BSC model with fixed $H(X|Y) = h(0.01)$, and $q = 0.15$.}
\label{f:data_FC_xx_01}
\end{figure}

In the forward BSC scenario, we focused on the case where the source is not u.d. (if it was u.d., then the model would be equivalent to the backward one analyzed above). In particular, we considered either that the probability of the source being one is $q=0.15$ or that it is $q=0.20$. As in the backward case, the decoder is not informed about this. In both cases, we assigned values of $p^f$ in order to obtain a target $H(X|Y)$. In particular, the targets that we employed are $h(0.10)$, $h(0.05)$, and $h(0.01)$, so that the expected optimal compression rates (but we know already that we will not operate at optimality) are equal to the ones expected in the previous section, relative to the backward BSC model. The experimental results are shown in Fig.~\ref{f:data_FC_xx_10}, Fig.~\ref{f:data_FC_xx_05}, and Fig.~\ref{f:data_FC_xx_01}, respectively for these three targets.

The plots compare the performance of the PD method, of the SSD method and of the \emph{quantized parity-based approach} (QPD method), and all are relative to a ``long'' frame. In the latter case, the lossy parity quantization is actually simulated by (i) adding to the parity bits a random binary noise having the same statistics of $Z^f$, (ii) assuming the rate is reduced by a factor $1 - H(Z^f)$. Note that, since at most the PD method permits to operate at $R_s = 2$ bit/sample, the maximum rate of the QPD method decreases to $[1 - H(Z^f)]R_s$, and it may be possible that no waterfall behavior could be seen, not even operating at this maximum rate. This is the reason why, for $q=0.15$ and $H(X|Y)=h(0.10)$ the QPD curve is not shown; similarly, since the SSD method is limited to $R_s = 2/3$ bit/sample, under the same settings no waterfall behavior could be seen, so the corresponding curve was not shown.

The theoretical limit found in section \ref{s:fw}, namely $H(Z^f)>H(X|Y)$, is also shown in the plots. Since in the case of $H(X|Y) = h(0.01)$ the theoretical losses for $q=0.15$ and $q=0.20$ are about the same (see Fig.~\ref{f:rateloss}), only the former case was investigated. The curve relative to the SSD method applied to a backward correlation model with \emph{same} conditional entropy (compression limit) and \emph{same} source distribution is shown too, in order to emphasize how different could be the experimental performances under different correlation models which may appear to be the same.

Similarly to the backward BSC case, we noticed that the SSD method is always better than the PD one, but we noticed also that the former has higher error-floor, especially for high correlation. However, the SSD method is always far from achieving the Slepian-Wolf bound, and also somewhat more far than expected from the theoretical limit $H(Z^f)$. Instead, if the model would be ``backward'' with same parameters, SSD would operate far way closer to the compression bound.

The differences between the $q=0.15$ and the $q=0.20$ case are very noticeable, suggesting that the performance should increase rapidly when approaching uniformity (see Fig.~\ref{f:rateloss}).

The QPD method, that in theory should be better than the PD method and operate under the same bound of the SSD method, did not provide the expected results. In particular, it improved the PD performance only for $H(X|Y) = h(0.10)$, while it always degraded with respect to the PD performance in the other cases. In practice, even if this possibility was not tested, it could be possible that QPD improves with respect to PD if operated with a slightly higher rate (i.e.~with less strong parity quantization).

\section{Conclusion}\label{s:conclusion}
In this paper, we reviewed the parity- and the syndrome- based approaches to the source coding problem with or without side information at the decoder. We discussed their theoretical limits, in particular in the case of a non-uniformly distributed source. Also, we recast the problem of decoding parities or syndromes formed with respect to turbo codes into a general maximum a posteriori probability problem. By using a factor-graph approach, we immediately devised how to take full advantage of the conventional iterative decoding algorithms traditionally employed in channel coding problems. We eventually used a unified perspective on the data reconstruction problem, that permits to deal straightforwardly with non-binary side information and with non-binary encoder-to-decoder transmission channels too.

Finally, we analyzed the performance of many different compression systems. The performance comparisons clearly show the differences between the parity- and the syndrome-based approaches, that are not usually discussed in the literature, in a variety of settings. Our experiments confirm the limits found in the theoretical analysis. The performance comparison with several other state-of-the art coding systems appeared in literature validates the practical utilization of the presented coding methods.

\IEEEtriggeratref{19}

\bibliographystyle{IEEEtran.bst}
\bibliography{IEEEabrv,nonIEEEabrv,refs_books,refs_DSC,refs_TCQ,refs_other,refs_my}

\end{document}

%% file: turbo_tcomm.txt
We consider the problem of compressing memoryless binary data with or without side information at the decoder. We review the parity- and the syndrome-based approaches and discuss their theoretical limits, assuming that there exists a virtual binary symmetric channel between the source and the side information, and that the source is not necessarily uniformly distributed. We take a factor-graph-based approach in order to devise how to take full advantage of the ready-available iterative decoding procedures when turbo codes are employed, in both a parity- or a syndrome-based fashion. We end up obtaining a unified decoder formulation that holds both for error-free and for error-prone encoder-to-decoder transmission over generic channels. To support the theoretical results, the different compression systems analyzed in the paper are also experimentally tested. They are compared against several different approaches proposed in literature and shown to be competitive in a variety of cases.